\documentclass{article}
\usepackage{spconf,amsmath,graphicx}
\usepackage{amssymb,amsmath,epsfig}
\usepackage{latexsym}
\usepackage{graphicx,multirow,threeparttable,float,balance,array,bm,subcaption,caption}
\usepackage{cite}
\usepackage{hyperref}
\usepackage[english]{babel}
\usepackage{diagbox}

\makeatletter
\renewcommand{\p@subfigure}{\thefigure.}
\makeatletter

\title{Time-domain speaker extraction network}
\name{Chenglin Xu$^{1,2}$, Wei Rao$^3$, Eng Siong Chng$^{1,2}$, Haizhou Li$^3$ \thanks{Published in ASRU 2019. This research is supported by Programmatic grant no. A1687b0033 from the Singapore government’s Research, Innovation and Enterprise 2020 plan (Advanced Manufacturing and Engineering domain). This research is supported by the National Research Foundation
Singapore under its AI Singapore Programme (and other co-funders, where
applicable). [AISG-100E-2018-006]}}
\address{
  $^1$ School of Computer Science and Engineering, Nanyang Technological University, Singapore \\
  $^2$ Temasek Laboratories@NTU, Nanyang Technological University, Singapore \\
  $^3$ Department of Electrical and Computer Engineering, National University of Singapore, Singapore\\
xuchenglin@ntu.edu.sg}
\begin{document}
%
\maketitle
\begin{abstract}
Speaker extraction is to extract a target speaker's voice from multi-talker speech. It simulates humans' cocktail party effect or the selective listening ability. The prior work mostly performs speaker extraction in frequency domain, then reconstructs the signal with some phase approximation. The inaccuracy of phase estimation is inherent to the frequency domain processing, that affects the quality of signal reconstruction. In this paper, we propose a time-domain speaker extraction network (TseNet) that doesn't decompose the speech signal into magnitude and phase spectrums, therefore, doesn't require phase estimation. The TseNet consists of a stack of dilated depthwise separable convolutional networks, that capture the long-range dependency of the speech signal with a manageable number of parameters. It is also conditioned on a reference voice from the target speaker, that is characterized by speaker i-vector, to perform the selective listening to the target speaker. Experiments show that the proposed TseNet achieves 16.3\% and 7.0\% relative improvements over the baseline in terms of signal-to-distortion ratio (SDR) and perceptual evaluation of speech quality (PESQ) under open evaluation condition.
\end{abstract}
%
\section{Introduction}
\label{sec:intro}

Human brain has the ability to focus the auditory attention on a particular voice while filtering out a range of other stimuli\cite{getzmann2017switching} in a cocktail party with multiple talkers. However, machines have yet to achieve the same attention ability as humans in the presence of background noise or interference of other speakers. Such auditory attention is required in real-world application, such as, speech recognition \cite{li2015robust,watanabe2017new,xiao2016study}, speaker verification\cite{rao2019target}, and speaker diarization\cite{sell2018diarization}. This paper proposes a novel technique for speaker extraction from a multiple talker speech.


The recent studies on speech separation and speaker extraction have advanced the state-of-the-art for solving the cocktails party problem. However, speech separation methods, such as Deep Clustering (DC) \cite{hershey2016deep,isik2016single,wang2018alternative}, Deep Attractor Network (DANet) \cite{chen2017deep}, Permutation Invariant Training (PIT) \cite{yu2017permutation,kolbaek2017multitalker,xu2018single,xu2018shifted}, Time-domain Audio Separation Net (TasNet) \cite{luo2018tasnet,luo2018real,luo2019conv}, require the number of speakers to be known in advance either during training or inference. For example, this prior information is needed in the PIT and TasNet methods during training, and in the DC and DANet methods during inference in order to form the exact number of speaker clusters. Such speech separation methods are greatly limited in real world applications as the number of speakers cannot always be known in practice.

Speaker extraction \cite{delcroix2018single,wang2018deep,xu2019optimization,xiao2019single,delcroix2019compact,ochiai2019unified} represents one of the solutions to overcome the problem of unknown speaker number. The extractor is conditioned on a reference voice from the target speaker to extract the voice of interest. This technique is particularly useful when the system is expected to respond only to the registered speakers where reference voices are available, for example, in speaker verification \cite{rao2019target,rao2019target_is}, and in the case where auditory attention should be given to a specific speaker. However, the prior works \cite{delcroix2018single,wang2018deep,xu2019optimization,xiao2019single,delcroix2019compact,ochiai2019unified} mostly perform speaker extraction in frequency domain, then reconstruct the signal with some phase approximation or simply the original phase from the mixture. The inaccuracy of phase estimation is inherent to the frequency domain processing, that affects the quality of signal reconstruction.


In this work, we propose a time-domain speaker extraction network (TseNet) that doesn't decompose the speech signal into magnitude and phase spectrums, therefore, doesn't require phase estimation. The TseNet is conditioned on the reference voice of the target speaker, that is characterized by speaker i-vector \cite{Dehak&Kenny2011}, to perform the selective listening to the target speaker. The prior information of number of speakers in the mixture is not needed any more, thus making the TseNet practical for many real-world applications. The TseNet consists of a stack of dilated depthwise separable convolutional networks, that captures the long-range dependency of the speech signal with a manageable number of parameters. It is optimized to maximize a scale-invariant signal-to-distortion ratio (SI-SDR) \cite{le2019sdr} loss.

\begin{figure*}[htb]
\hspace*{\fill}
\begin{subfigure}[b]{0.7\textwidth}
  \centering
  \includegraphics[width=\textwidth]{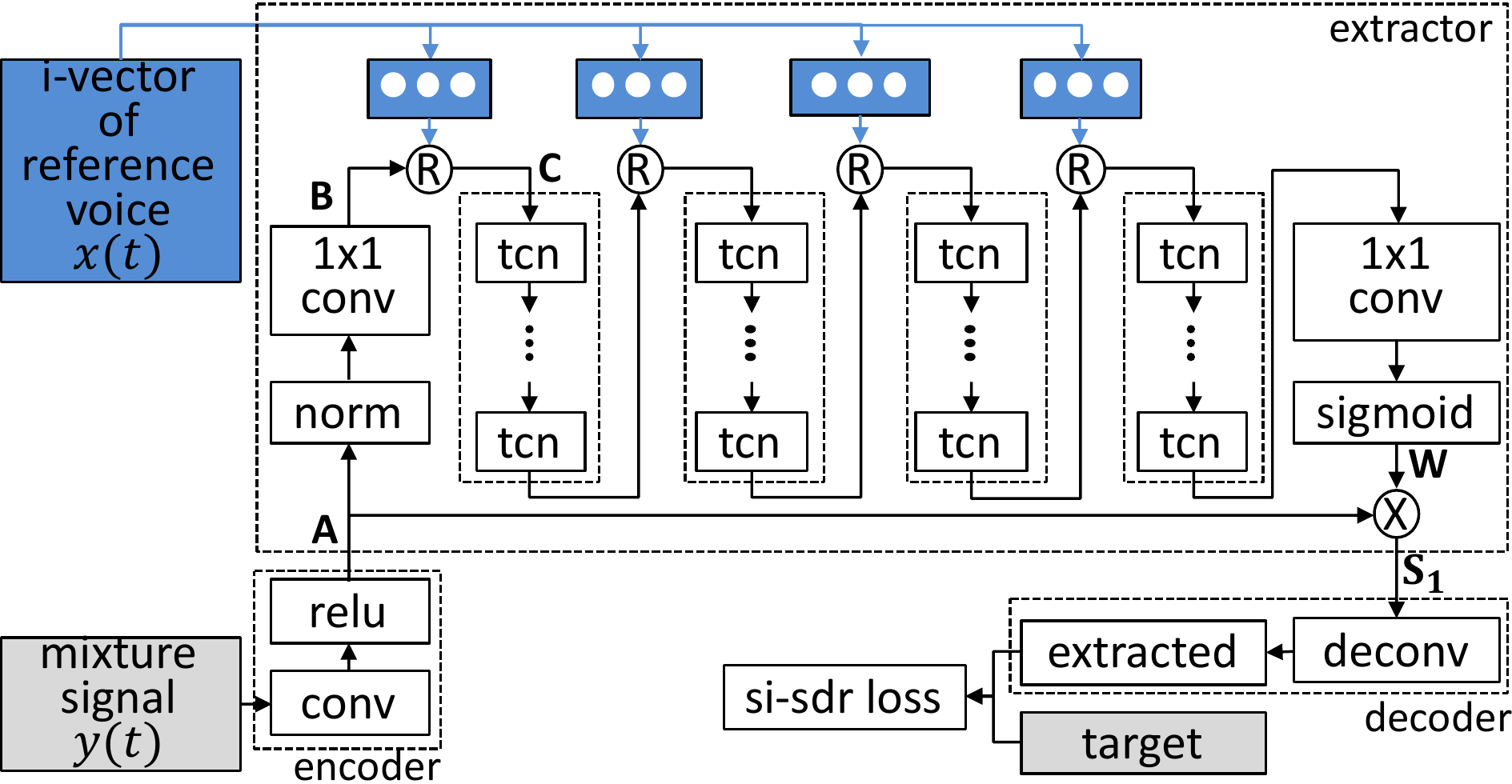}\medskip
  \caption{TseNet.}
  \label{fig:tsenet}
\end{subfigure}
\hfill
\begin{subfigure}[b]{0.15\textwidth}
  \centering
  \includegraphics[width=0.8\textwidth,scale=0.7]{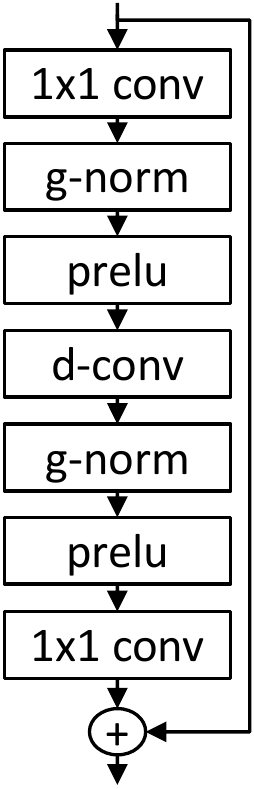}\medskip
  \caption{TCN block.}
    \label{fig:tcn}
\end{subfigure}
\hspace*{\fill}
\caption{The framework of the proposed TseNet approach is shown in Fig.\ref{fig:tsenet}. \textcircled{\raisebox{-0.9pt}{R}} is an operator that concatenates the speaker vector repeatedly to the intermediate representations of mixture speech along the channel dimension. \textcircled{\raisebox{-0.9pt}{X}} refers to the element-wise multiplication. The ``conv'' and ``deconv'' are 1-D convolutional and de-convolutional operations. ``relu'' and ``sigmoid'' are the rectified linear unit (ReLU) and sigmoid functions. The top four blocks with i-vector as inputs are dense layers with ReLU activation function. The structure of the ``tcn'' block is temporal convolutional network as shown in Fig.\ref{fig:tcn}. \textcircled{\raisebox{-0.9pt}{+}} denotes the residual connection. The ``d-conv'' is depthwise convolution which forms a depthwise separable convolution together with the last ``1x1 conv''. ``prelu'' is the parametric rectified linear unit (PReLU). ``g-norm' is the mean and variance on both dimensions of time frames and channels scaled by the trainable bias and gain parameters.}
\vspace{-10pt}
\label{fig:sys}
\end{figure*}

The rest of the paper is organized as follows. The proposed TseNet is detailed in Section \ref{sec:system}. Section \ref{sec:exp} reports the experimental setup and results. Section \ref{sec:con} concludes the study.

\vspace{-5pt}
\section{Time-domain Speaker Extraction Network}
\label{sec:system}
\vspace{-5pt}

In time-domain speaker extraction, we suppose that a signal $y(t)$ of $T$ samples is the mixture of the target speaker's voice $s_1(t)$ and $I-1$ interference voices or background noise $s_i(t)$, as formulated next,
\begin{equation} \label{eq:discret_signal}
y(t) = s_1(t) + \sum_{i=2}^I s_i(t), \quad t=1,...,T
\end{equation}
where $I$ might be any number of interferences, and $s_i(t)$ might be either interference speech or background noise.

During the inference at run-time, given a mixture signal $y(t)$ and a reference voice $x(t)$, the speaker extractor is expected to estimate $\hat{s}_1(t)$ that is close to $s_1(t)$ subject to an optimization criterion.


\vspace{-5pt}
\subsection{TseNet Architecture}
\label{ssec:tsenet}
\vspace{-5pt}

We propose a time-domain speaker extraction network to estimate the target speaker's voice $\hat{s}_1(t)$ from the mixture speech $y(t)$. The TseNet is conditioned on a reference voice $x(t)$ of the target speaker that is characterized by a speaker i-vector \cite{Dehak&Kenny2011}. I-vector is one of the most used techniques to characterize a speaker.

Fig.\ref{fig:sys} shows the framework of the proposed TseNet, which consists of three components: an encoder, an extractor and a decoder. The encoder transforms the time-domain mixture signal into a representation using a convolutional neural network (CNN). Such representation is then taken as input to the extractor. Specifically, the representation is used to estimate a mask for the target speaker at each time step together with the target speaker's i-vector, which represents the target speaker. After that, only the target speaker's representation is extracted by filtering the encoded representation of the input mixture with the estimated mask. Finally, the time-domain signal of the target speaker is reconstructed from the extracted representation of the target speaker with a decoder. The details of the three components are described as follows.

\vspace{-5pt}
\subsubsection{Encoder}
\label{sssec:encoder}
\vspace{-5pt}

The input mixture speech $y(t)\in\mathbb{R}^{1\times T}$ is encoded to a representation $A\in \mathbb{R}^{K\times M}$ by a 1-D CNN with $M$ filters and a filter size of $L$ samples with a stride of $L/2$ samples followed by a rectified linear unit (ReLU) activation function. Each vector $A_k$ of the representation $A$ is defined as,
\begin{equation}
    A_k = \text{ReLU}(y_k*U), \quad k=1,...,K
\end{equation}
where $K=2(T-L)/L+1$, and $y_k\in \mathbb{R}^{1\times L}$ is the $k^{th}$ segment of $y(t)$, which is divided by every $L$ samples with an overlapping of $L/2$ samples. $U\in \mathbb{R}^{M\times L}$ is the encoder basis.

\subsubsection{Extractor}

The extractor as shown in Fig.\ref{fig:tsenet} is designed to obtain the extracted representation of the target speaker $S_1$ by estimating a mask $W$ for the speaker extraction from the encoded representations $A$. It can be formulated as
\begin{equation} \label{Eq:s1}
    S_1 = W \otimes A
\end{equation}
where $\otimes$ is an operator for element-wise multiplication. 

Specifically, the encoded representation $A$ is firstly normalized by its mean and variance on channel dimension scaled by the trainable bias and gain parameters. Then, a $1\times 1$ CNN with $N$ filters is performed to adjust the number of channels for the inputs and residual path of the subsequent blocks of temporal convolutional network (TCN). The encoded representation becomes $B\in \mathbb{R}^{K\times N}$. At the same time, a dense layer with ReLU activation function is introduced to take the speaker i-vector $I_1\in \mathbb{R}^{1\times D_1}$, and its outputs $I_2\in \mathbb{R}^{1\times D_2}$ are repeated and concatenated with the encoded representation $B$. The encoded representation will contain speaker information as $C\in \mathbb{R}^{K\times (D_2+N)}$. A non-linear dense layer is always added to fine tune the speaker i-vector before feeding it into the subsequent TCN blocks.

To capture the long-range dependency of the speech signal with a manageable number of parameters, dilated depthwise separable convolution networks are stacked in several TCN blocks by exponentially increasing the dilation factor. Each TCN block, as shown in Fig. \ref{fig:tcn}, consists of two $1\times 1$ CNNs to determine the number of input and output channels of the depthwise convolution. The first $1\times 1$ CNN has $O$ filters with a $1\times 1$ kernel size. The shape of the filters in the depthwise convolution will be $[1\times P, O]$, where $1\times P$ is the kernel size. The outputs from the depthwise convolution have a channel size of $O$. The second $1\times 1$ CNN with $N$ filters adjust the channel size of the outputs of the depthwise convolution from $O$ to $N$. The inputs with $N$ channels of the TCN block are then added to its outputs with $N$ channels by a residual path, except the TCN block with the concatenated representations $C$ as its inputs, where the encoded representation before the concatenation will be added to the outputs of this TCN block. We form $b$ TCN blocks as a batch and repeat the batch for $r$ times in the extractor. In each batch, the dilation factors of the depthwise convolutions in the $b$ TCN blocks will be increased as $[2^0,...,2^{(b-1)}]$.


Because the estimated mask $W$ will be used for the extraction from the encoded representations $A$, the dimensions of $W$ should be same as $A$. Since the output channels $N$ from the last TCN block may be different from the channels $M$ of the encoded representations $A\in\mathbb{R}^{K\times M}$, we would like to apply one $1\times 1$ CNN to transform the dimension of the output channels from the last TCN block to be same as the encoded representations $A\in \mathbb{R}^{K\times M}$. Therefore, the elements of the mask $W\in \mathbb{R}^{K\times M}$ are estimated through a sigmoid activation function to ensure that they range within $[0,1]$. Finally, the extracted representation of the target speaker, $S_1\in \mathbb{R}^{K\times M}$, is estimated by Eq. \ref{Eq:s1}.

\subsubsection{Decoder}
\label{sssec:decoder}

The time-domain target speaker's voice $\hat{s}_1$ is reconstructed from the extracted representation $S_1$ by a decoder, which is a de-convolutional operation.
\begin{equation}
    \hat{s}_1 = S_1*V
\end{equation}
where $V\in \mathbb{R}^{M\times L}$ is the decoder basis.

\subsection{Training Objective}
\label{ssec:objective}

The TseNet is optimized by maximizing the scale-invariant signal-to-distortion ratio (SI-SDR) \cite{le2019sdr}. The SI-SDR is defined as:
\begin{equation}
    \text{SI-SDR} = 10\log_{10}(\frac{||\frac{\langle\hat{s}, s\rangle}{\langle s,s\rangle}s||^2}{||\frac{\langle\hat{s}, s\rangle}{\langle s,s\rangle}s-\hat{s}||^2})
\end{equation}
where $\hat{s}$ and $s$ are the extracted and target signals of the target speaker, respectively. $\langle\cdot,\cdot\rangle$ is the inner product. To ensure scale invariance, the signals $\hat{s}$ and $s$ are normalized to zero-mean prior to the SI-SDR calculation. The SI-SDR calculation between the extracted and target signals in Fig.\ref{fig:tsenet} is performed only during training, and is not required for run-time inference.

\section{Experiments and Discussion}\vspace{-5pt}
\label{sec:exp}

\subsection{Data}
\label{subsec:data}

A two speakers mixture database\footnote{The database simulation code is available at: \url{https://github.com/xuchenglin28/speaker_extraction}} was generated at sampling rate of 8kHz based on the WSJ0 corpus \cite{garofolo1993csr}. We simulated three datasets: training set ($20,000$ utterances), development set ($5,000$ utterances), and test set ($3,000$ utterances). In particular, the training set and development set were generated by randomly selecting two utterances from $50$ male and $51$ female speakers in the WSJ0 ``si\_tr\_s'' set at various signal-to-noise ratio (SNR) between $0$dB and $5$dB. Similarly, the utterances from $10$ male and $8$ female speakers in the WSJ0 ``si\_dt\_05'' set and ``si\_et\_05'' set were randomly selected to create the mixed test set. Since the speakers in the test set were excluded from the training and development sets, the test set was used to evaluate the speaker independent performance and regarded as open condition evaluation.

To include both overlapping and non-overlapping speech in the dataset, we used the longer duration of the two utterances as the duration of the mixed speech when we simulated the mixture database \footnote{In most source separation approaches based on WSJ0 corpus \cite{hershey2016deep,isik2016single,wang2018alternative,chen2017deep,yu2017permutation,kolbaek2017multitalker,xu2018single,xu2018shifted,luo2018tasnet,luo2018real,luo2019conv}, the minimum length of the two utterances was kept as the length of the mixture.}. The speaker of the first randomly selected utterance was regarded as the target speaker, the second utterance of a different speaker was randomly selected as interference speech. The first selected utterance from the original WSJ0 corpus was used as the target signal to supervise the optimization of the network during training. At run-time, the speaker extraction process was conditioned on a reference signal from the target speaker, which was different from the one of the target speaker in the mixture, and was randomly selected.

\subsection{Experimental Setup}
\label{subsec:setup}

\subsubsection{I-vector extractor}

An i-vector extractor, that included a universal background model (UBM) and total variability matrix, was developed to convert a speech utterance into a low dimension vector representation. We first trained the UBM and total variability matrix with the single speaker (clean) speech from the training and development sets. The extractor was trained with the features of 19 MFCCs together with energy plus their 1st- and 2nd-derivatives extracted from the speech regions, followed by cepstral mean normalization \cite{Atal74} with a window size of 3 seconds. The 60-dimensional acoustic features were extracted from a window length of 25ms with a shift of 10ms. A Hamming window was applied. An energy based voice activity detection method was used to remove the silence frames. The i-vector extractor was based on a gender-independent UBM with 512 mixtures and a total variability matrix with 400 total factors.
\vspace{-8pt}
\subsubsection{Network configuration}

The speaker i-vector had a dimension of $D_1(=400)$ in this work. As the i-vector extractor was not jointly trained with the TseNet, we applied dense layers with ReLU activation function to fine tune the i-vector prior to each concatenation in the extractor part of TseNet. After the non-linear dense layer, the dimension of the i-vector became $D_2(=100)$.

The mixture input $Y\in\mathbb{R}^{1\times T}$ was encoded by a 1d-convolution with $M(=256)$ filters followed by a rectified linear unit (Relu) activation function. Each filter had a window of $L(=20)$ samples and a stride of $L/2(=10)$ samples. 

In the extractor part of the TseNet, a mean and variance normalization with trainable gain and bias parameters was applied to the encoded representations $A\in\mathbb{R}^{K\times M}$ on the channel dimension, where $K$ was equal to $2(T-L)/L+1$. A 1x1 convolution linearly transformed the normalized encoded representations $A$ to the representations $B\in\mathbb{R}^{K\times N}$ with $N(=256)$ channels, which determined the number of channels in the input and residual path of the subsequent $1\times 1$ CNN. The number of input channels $O$ and the kernel size $1\times P$ of each depthwise convolution were set to $512$ and $1\times 3$. $b(=8)$ TCN blocks were formed as a batch and repeated for $r(=4)$ times.

The network was optimized by the Adam algorithm \cite{kingma2014adam}. The learning rate started from $0.001$ and was halved when the loss increased on the development set for at least $3$ epochs. Early stopping scheme was applied when the loss increased on the development set for $10$ epochs. The utterances in the training and development set were broken into $4$s segments, and the minibatch size was set to $10$.
\vspace{-8pt}
\subsubsection{Baselines}
\label{sec:baselines}

We selected $4$ baselines that represented the recent advancements in single channel target speaker extraction. We first implemented the initial work using speaker beam as a front-end (SBF) \cite{delcroix2018single} with two different loss functions as two baselines. Bidirectional long short term memory network (BLSTM) was applied in the speaker extraction network followed by a speaker adaptation layer to include target speaker's information. The other two baselines \cite{xu2019optimization} by extending the SBF methods were also implemented. BLSTM was applied in both speaker extraction network and speaker embedding network for target speaker's information to capture long temporal information. With the temporal constraints added in the loss function, the baselines in \cite{xu2019optimization} further advanced the performance achieved in \cite{delcroix2018single}.
\begin{itemize}
    \item SBF-IBM \cite{delcroix2018single}: applied a speaker adaptation layer in a context adaptive deep neural network (CADNN) \cite{delcroix2016context} to track the target speaker from the input mixture. The adaptation weights were jointly learned from a target speaker's enrolled speech. A mask approximation loss was calculated between the estimated mask and ideal binary mask (IBM) as the objective to optimize the network. \vspace{-5pt}
    \item SBF-MSAL \cite{delcroix2018single}: replaced the mask approximation loss in the SBF-IBM method to a magnitude spectrum approximation loss (MSAL) to calculate direct signal reconstruction error. \vspace{-5pt}
    \item SBF-MTSAL \cite{xu2019optimization}: introduced a magnitude and temporal spectrum approximation loss (MTSAL) into the SBF-MSAL by adding temporal loss as weighted constraints. \vspace{-5pt}
    \item SBF-MTSAL-Concat \cite{xu2019optimization}: changed the speaker embedding network to obtain speaker characteristics by a BLSTM to capture long temporal information. Instead of the speaker adaptation layer in the CADNN structure, the speaker embedding was repeated and concatenated to each frame of the representations, that encoded the input mixture. \vspace{-5pt}
\end{itemize}
\vspace{-5pt}
\subsubsection{Evaluation metrics}
\vspace{-5pt}
We evaluated the performance with the criteria of signal-to-distortion ratio (SDR) \cite{vincent2006performance}, SI-SDR \cite{le2019sdr},  and perceptual evaluation of speech quality (PESQ) \cite{rix2001perceptual}. The subjective evaluation of A/B preference test was also conducted.

\subsection{Results}
\label{subsec:results}
\vspace{-5pt}
\subsubsection{Overall comparisons}
\vspace{-5pt}
Table \ref{tbl:methods_comp} summarizes the SDR, SI-SDR and PESQ performances of the SBF based baselines \cite{delcroix2018single,xu2019optimization} and the proposed TseNet method. Compared with the best baseline, SBF-MTSAL-Concat, the proposed TseNet approach achieves $16.3\%$, $15.4\%$ and $7.0\%$ relative improvements in terms of SDR, SI-SDR and PESQ with approximate same number of parameters under open condition, where both the target and interference speakers are unseen by the system.

It is noted that TseNet significantly outperforms the baselines system. The time-domain processing has shown its clear advantage over its frequency-domain counterparts. Because the TseNet doesn't decompose the speech signal into magnitude and phase spectrums, therefore, doesn't require phase estimation. In addition, the TseNet also benefits from the long-range dependency of the speech signal captured by the stacked dilated depthwise separable convolution with a manageable number of parameters. Without the recurrent connection, the TseNet method can be easily parallelized for fast training and inference.

\begin{table}[t]
\centering \caption{ SDR (dB), SI-SDR(dB) and PESQ in a comparative study of different techniques under open condition. ``Mixture'' refers to original input mixture with zero effort. ``\#Paras'' means the number of parameters of the model.} 
\centerline{
\footnotesize
\begin{tabular}{|r|*{5}{c|}}
\hline
Methods & \#Paras & SDR & SI-SDR  & PESQ  \\
\hline
\hline
Mixture & - & 2.60 & 2.51 & 2.31 \\
\hline
SBF-IBM \cite{delcroix2018single} & 19.3M & 6.45 & 6.27 & 2.32 \\ 
SBF-MSAL \cite{delcroix2018single} & 19.3M & 9.62 & 9.22 & 2.64 \\
SBF-MTSAL \cite{xu2019optimization} & 19.3M & 9.90 & 9.49 & 2.66 \\ 
SBF-MTSAL-Concat \cite{xu2019optimization} & 8.9M & 10.99 & 10.56 & 2.73 \\ \hline
TseNet & 9.0M & \textbf{12.78} & \textbf{12.19} & \textbf{2.92} \\
\hline
\end{tabular}} 
\label{tbl:methods_comp}
\end{table}

\begin{figure}[t]
\begin{center}
\includegraphics[width=80mm]{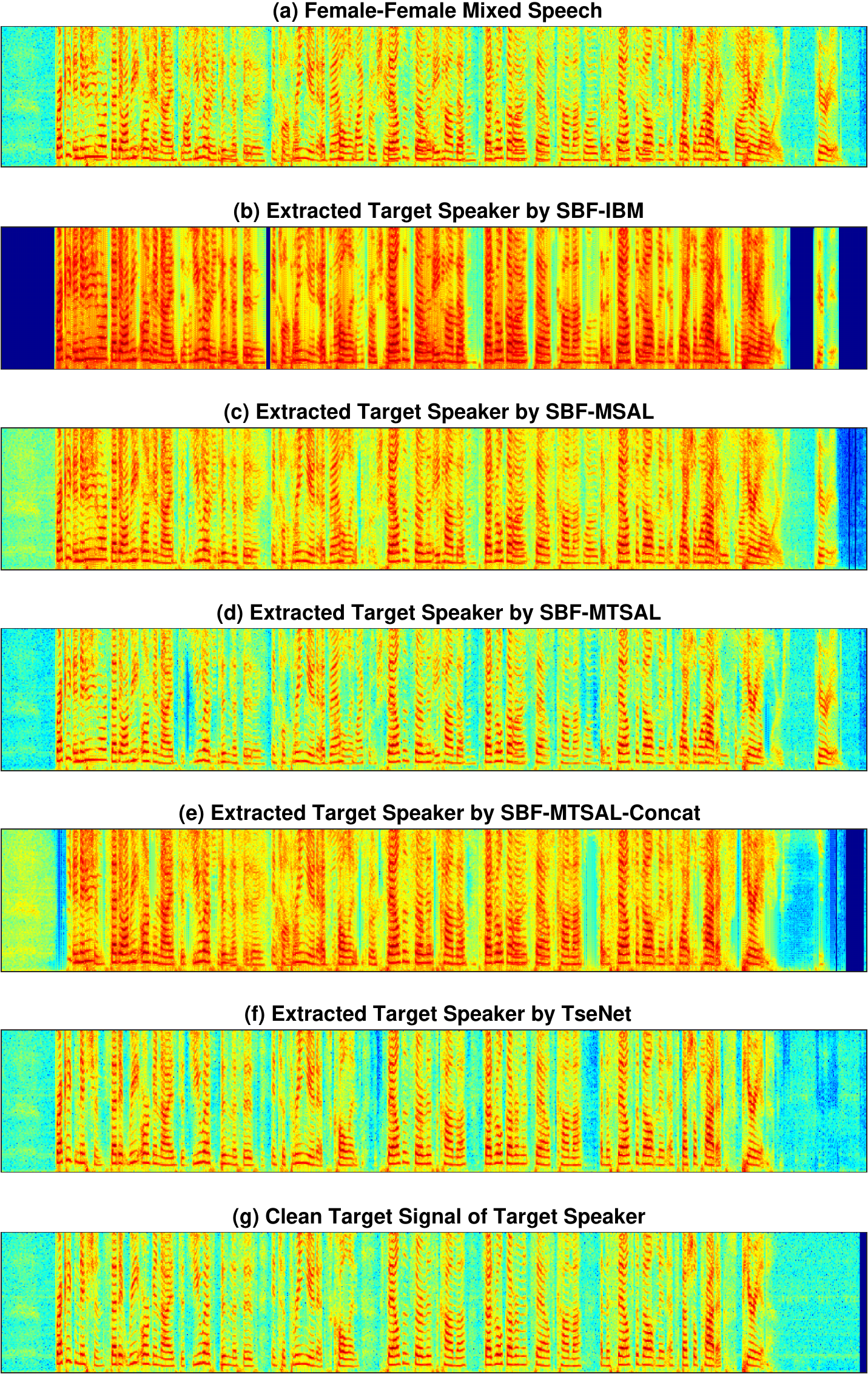} 
\caption{The log magnitude spectrum illustrations of a female-female mixture, extracted target speaker's signals by the baselines and the proposed TseNet, and the clean target signal.}
\label{fig:spec}
\end{center}
\vspace{-25pt}
\end{figure}

One female-female mixture example is selected to illustrate the differences of extracted target speaker's voices between the baselines and the proposed TseNet methods, as shown in Fig. \ref{fig:spec}. From the log magnitude spectrum, we observe that the proposed TseNet outperforms other baseline methods in terms of the recovered signal quality and purification. Some listening examples are available online \footnote{\url{https://xuchenglin28.github.io/files/asru2019/index.html}. The first example on the web page is the audio illustrated in Fig. \ref{fig:spec}.}.

\subsubsection{Different vs. same gender}

We further analysis and summarize the performances for different and same gender mixture in Table \ref{tbl:gender}. We observe that the performance on different gender mixture is always better than the same gender. Since the characteristics, i.e., pitch, of the same gender speakers are much closer than the speakers of different gender, the target speaker extraction task becomes difficult and hard when the interference speaker's gender is same as the target speaker. From Table \ref{tbl:gender}, we observe that the proposed TseNet method achieves $20.8\%$ and $8.7\%$ relative improvement over the best baseline system, SBF-MTSAL-Concat, on SDR for different and same gender conditions. At the same time, the PESQs for different and same gender conditions relatively improve $9.3\%$ and $3.5\%$ over the best baseline system, SBF-MTSAL-Concat.

\begin{table}[t] 
\centering \caption{SDR (dB) and PESQ in a comparative study of different and same gender mixture under open condition.} 
\centerline{
\small
\begin{tabular}{|r|*{4}{c|}}
\hline
\multirow{2}{*}{Methods} & \multicolumn{2}{c|}{SDR} & \multicolumn{2}{c|}{PESQ} \\ \cline{2-5}
 & Diff. & Same & Diff.  & Same  \\
\hline
\hline
Mixture & 2.51 & 2.69 & 2.29 & 2.34 \\
\hline
SBF-IBM \cite{delcroix2018single} & 7.61 & 5.13 & 2.42 & 2.19 \\ 
SBF-MSAL \cite{delcroix2018single} & 12.01 & 6.87 & 2.82 & 2.43 \\ 
SBF-MTSAL \cite{xu2019optimization} & 12.27 & 7.17 & 2.85 & 2.44 \\ 
SBF-MTSAL-Concat \cite{xu2019optimization} & 12.87 & 8.84 & 2.90 & 2.54 \\ \hline
TseNet & \textbf{15.55} & \textbf{9.61} & \textbf{3.17} & \textbf{2.63} \\
\hline
\end{tabular}}
\label{tbl:gender} 
\end{table}

\vspace{-2.5pt}
\subsubsection{Mixture with different SNR}
\vspace{-2.5pt}

It is of interest to investigate how the proposed TseNet performs for mixture speech of different SNR, where we consider the target speech as the signal and the interference as the noise. We divide the test set into $3$ subsets by the SNR range of [0, 1)dB, [1, 3)dB and [3, 5]dB. The results are summarized in Table \ref{tbl:diff_snr} according to the SNR range. We observe that the speaker extraction performs better with speech mixture of higher SNR. We also observe that the proposed TseNet method achieves $30.8\%$, $18.9\%$ and $9.1\%$ relative improvement over the best baseline system, SBF-MTSAL-Concat, on SDR for different SNRs of [0, 1)dB, [1, 3)dB and [3, 5]dB, respectively. We are glad to see TseNet works well across different SNR levels.

\begin{table}[t] 
\centering \caption{SDR (dB) in a comparative study of the speaker extraction performance of the mixture with different SNR, which is divided into [0, 1)dB, [1, 3)dB, [3, 5]dB.} 
\centerline{
\small
\begin{tabular}{|r|*{3}{c|}}
\hline
\diagbox{Methods}{SNR(dB)} & [0, 1) & [1, 3) & [3, 5]  \\
\hline
\hline
Mixture & 0.72 & 1.98 & 4.16 \\
\hline
SBF-IBM \cite{delcroix2018single} & 3.96 & 5.82 & 8.37 \\ 
SBF-MSAL \cite{delcroix2018single} & 7.11 & 9.23 & 11.27 \\ 
SBF-MTSAL \cite{xu2019optimization} & 7.48 & 9.50 & 11.52 \\ 
SBF-MTSAL-Concat \cite{xu2019optimization} & 8.67 & 10.62 & 12.52 \\ \hline
TseNet & \textbf{11.34} & \textbf{12.63} & \textbf{13.66} \\ 
\hline
\end{tabular}}
\label{tbl:diff_snr} 
\end{table}

%

\vspace{-8pt}
\subsubsection{Subjective evaluation}
\vspace{-5pt}

Since the SBF-MTSAL-Concat represents the best baseline performance in the objective evaluation, we only conduct an A/B preference test on the extracted target speaker's voices between the proposed TseNet and the best SBF-MTSAL-Concat baseline to evaluate the signal quality and intelligibility for listening. We randomly select $20$ pairs of listening examples, including the original target speaker's reference and two extracted signals for the target speaker by the proposed TseNet and the best baseline method. We invite a group of $12$ subjects to select their preference according to the quality and intelligibility. The listeners pay special attention to the amount of perceived interference from background voices. When the subject listens to each pair of the three audios, the reference signal is firstly played, and other two signals are randomly played one-by-one. The subject doesn't know which method the second and third audios are extracted by. 

\begin{figure}[!t]
\begin{center}
\includegraphics[width=85mm]{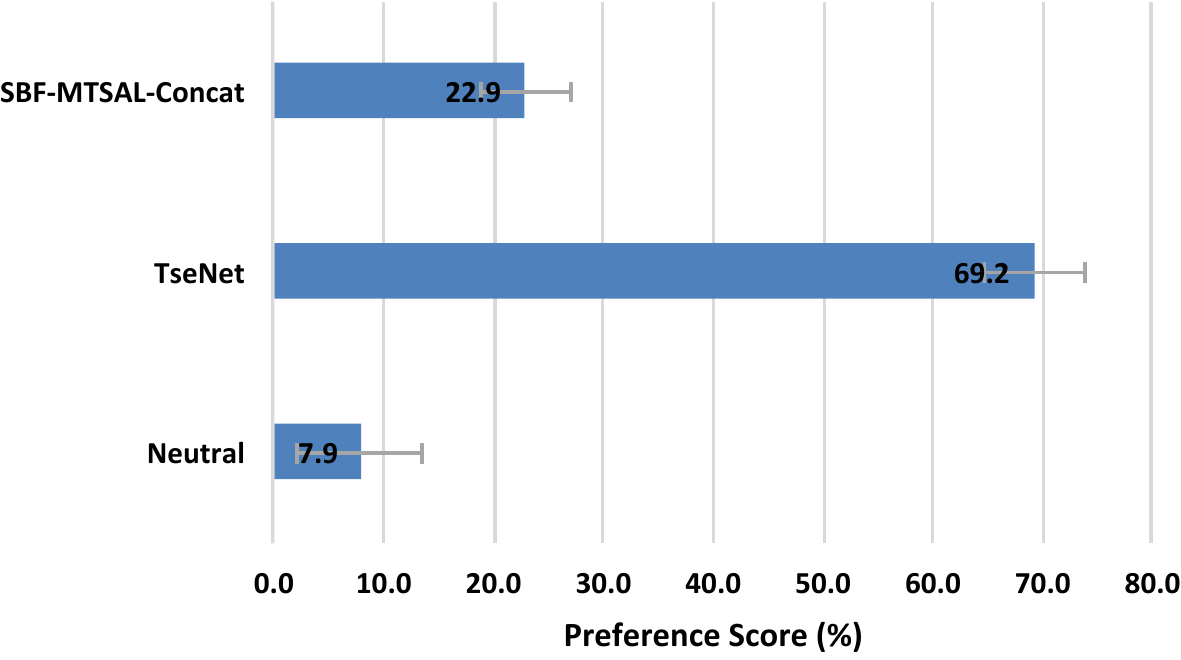}
\caption{The A/B preference test result of the extracted target speaker's voice between the proposed TseNet method and the best SBF-MTSAL-Concat baseline. We conducted t-test using a significance level of $p<0.05$ which is depicted with the error bars.}
\label{fig:ABtest}
\end{center}
\vspace{-25pt}
\end{figure}

The percentage of the preferences is shown in Fig. \ref{fig:ABtest}. We observe that the listeners clearly prefer the proposed TseNet method with a preference score of $69.2\%$ to the best SBF-MTSAL-Concat baseline with a preference score of $22.9\%$. Most subjects significantly prefer the extracted audios by our TseNet method with a significance level of $p<0.05$, because there are less distortion and inter-speaker interference than the best baseline.

\vspace{-8pt}
\subsection{Discussion}
\vspace{-5pt}

In this work, the speaker i-vector is used to model the characteristics of the target speaker. Although the i-vector is pre-computed for each speaker, we have designed several non-linear dense layers to fine tune the i-vector. These layers are jointly trained with the main speaker extraction network. In addition, the i-vector extractor system may need less data to be well trained than the neural network based speaker embedding system, i.e., x-vector system \cite{snyder2016deep}. Although the neural network based speaker embedding system can be jointly trained with the main speaker extraction network from scratch, it only works well when a huge of data is available for the network training. In the future, we will explore the effectiveness of the jointly trained speaker embedding network instead of i-vector system.

\vspace{-10pt}
\section{Conclusions}
\label{sec:con}
\vspace{-10pt}

In this paper, we proposed a time-domain speaker extraction network (TseNet) to avoid the phase estimation in the frequency-domain methods. The TseNet also doesn't need the prior information of the number of speakers in the mixture speech, therefore, becoming a practical solution to speaker extraction. Experimental results show that the proposed TseNet method outperforms other frequency-domain baseline methods significantly in terms of SDR and PESQ. The subjective test also shows that the proposed TseNet method is significantly preferred comparing with the best SBF-MTSAL-Concat baseline.


\bibliographystyle{IEEEbib}
\bibliography{asru2019}

\end{document}